\documentclass[preprint]{aastex}
\usepackage{color}
\newcommand{\kms}{{km~s$^{-1}$~}}
\newcommand{\feii}{{Fe~{\sc ii}~}}
\newcommand{\civ}{{C~{\sc iv}~}}
\newcommand{\ciii}{{C~{\sc iii}}}
\newcommand{\nv}{{N~{\sc v}~}}
\newcommand{\mgii}{{Mg~{\sc ii}~}}
\newcommand{\heii}{{He~{\sc ii}~}}
\newcommand{\aliii}{{Al~{\sc iii}~}}

\newcommand{\oiii}{{O~{\sc iii}}}

\usepackage{amsmath}  
\numberwithin{equation}{subsection}

\newcommand*{\appheading}[1][Appendix]{%
  \setcounter{secnumdepth}{0}\section{#1}\setcounter{secnumdepth}{3}%
  \renewcommand*{\thesubsection}{\Alph{subsection}}
  \numberwithin{equation}{subsection}
  \numberwithin{figure}{subsection}
}

\begin{document}

\title{Evidence for Fluorescent Fe~{\sc II} Emission from Extended Low 
Ionization Outflows in Obscured Quasars} 
\author{Tinggui Wang\altaffilmark{1}, Gary J. Ferland\altaffilmark{2}, Chenwei Yang\altaffilmark{1}, Huiyuan Wang\altaffilmark{1} and Shaohua Zhang\altaffilmark{3}}
\altaffiltext{1}{
CAS Key Laboratory for Researches in Galaxies and Cosmology, 
Department of Astronomy, University of Science and Technology of China, 
Hefei, Anhui 230026; twang@ustc.edu.cn
}
\altaffiltext{2}{Department of Physics and Astronomy, University of Kentucky, Lexington, KY 40506, USA}
\altaffiltext{3}{Polar Research Institute of China, 451 Jinqiao Rd, Shanghai 200136, China}

\begin{abstract}

Recent studies have shown that outflows in at least some broad absorption line (BAL) 
quasars are extended well beyond the putative dusty torus. Such outflows should be 
detectable in obscured quasars. We present four WISE selected infrared red quasars 
with very strong and peculiar ultraviolet Fe\,{\sc ii} emission lines: strong UV 
Fe\,{\sc ii} UV arising from transitions to ground/low excitation levels, and very 
weak Fe\,{\sc ii} at wavelengths longer than 2800\AA. The spectra of these quasars 
display strong resonant emission lines, such as C\,{\sc iv}, Al\,{\sc iii} and 
Mg\,{\sc ii} but sometimes, a lack of non-resonant lines such as C\,{\sc iii}], 
S\,{\sc iii}  and He\,{\sc ii}. We interpret the Fe\,{\sc ii} lines as resonantly 
scattered light from the extended outflows that are viewed nearly edge-on, so that 
the accretion disk and broad line region are obscured by the dusty torus, while 
the extended outflows are not.  We show that dust free gas exposed to strong radiation 
longward of 912\AA~ produces Fe\,{\sc ii} emission very similar to that observed. 
The gas is too cool to collisionally excite Fe\,{\sc ii} lines, accounting for the 
lack of optical emission. The spectral energy distribution from the UV to the 
mid-infrared can be modeled as emission from a clumpy dusty torus, with UV emission 
being reflected/scattered light either by the dusty torus or the outflow. Within 
this scenario, we estimate a minimum covering factor of  the outflows from a few 
to 20 percent for the Fe\,{\sc ii} scattering region, suggesting that Fe\,{\sc ii} 
BAL quasars are at a special stage of quasar evolution.

\end{abstract}


\section{Introduction}

Outflows are ubiquitous among quasars. They manifest themselves as 
blue-shifted broad and narrow absorption lines in UV and X-rays (e.g., 
Boksenberg et al. 1978; Turnshek et al. 1980; Weyman et al. 1991). The 
broad absorption lines (BALs) with widths of around a few thousands to 
a few ten thousands \kms~appear in 15-25\% of all quasars (e.g.,Weymann 
et al. 1991; Knigge et al. 2008), and intrinsic narrow absorption lines 
(NALs) with widths around hundreds to one thousand \kms~in 40-50\% 
of quasars (Nester et al. 2008). Almost all known BAL QSOs show broad high 
ionization absorption lines, such as \civ, \nv (Weymann et al. 1991; cf, 
Mrk 231, Veilleux et al. 2013), while only a small fraction (about 25\%) 
of BAL QSOs display also low ionization broad absorption lines (LoBALs, 
Weymann et al. 1991; Zhang et al. 2010; c.f. Urrutia et al. 2009), commonly 
\mgii. A yet smaller subset of LoBAL QSOs ($<1$\% of all quasars) exhibit 
broad \feii~absorption lines (hereafter, FeLoBALs; Trump et al. 2006; 
c.f., Dai et al. 2012). 

There are two different scenarios for BAL and non-BAL quasars or for a 
subclass of BAL QSOs. In the first scenario, the presence or the absence 
of BALs, or more generally, the manifestation of different subclasses of 
BALs, are attributed to the different viewing angles of the  same intrinsic 
population (Weymann et al. 1991). While in the second scenario, LoBAL and 
non LoBAL QSOs are at different evolutionary stages (Boroson \& Meyers 
1992; Voit et al. 1993; Hall et al. 2002). In other words, only a small 
fraction of quasars possess low-ionization outflows. Early studies showed 
that the emission lines and broad-band spectral energy distributions 
(SED) are very similar for HiBAL and non BAL quasars (Weymann et al. 1991), 
suggesting the unification scheme. Hamann, Korista \& Morris (1993) 
constrained the global covering factor to less 20\% from the fraction of 
resonant scattering light. However, systematic trends in the SED and 
emission lines have been revealed with a large sample of BAL, LoBAL in 
particular, and a non-BAL quasar sample from SDSS (e.g., Zhang et al. 2010; 
Baskin, Laor \& Hamann 2013; Farrah et al. 2010; Wang et al. 2013; and 
reference therein; c.f., Lazarova et al. 2012), suggesting that the global 
covering factor and properties of LoBALs depend on the intrinsic properties 
of the system. 
  
Feedback through these outflows is now widely believed to be a key physical 
process in the formation of massive galaxies (Fabian 2012). However, the 
mass loss rate and kinetic power, as well as the size, of outflows are 
still poorly constrained for the majority of BAL quasars (Lucy et al. 2014). 
To estimate these parameters from absorption lines, one needs to know the 
global covering factor, column density and the gas density or distance. The gas 
column density can be inferred from the column densities of a set of ions 
through photoionization modeling. However, the gas density diagnostics for 
absorption lines are available for only a rare subset of BAL QSOs with 
measured absorption lines from low-lying excitation levels of heavy elements 
(de Kool et al. 2001; Korista et al. 2008; Dunn et al. 2012). Previous 
measurements of density in a dozen of BALs, mostly FeLoBALs, quasars, indicate 
that BAL region (BALR) is extended from 100 parsec to several kpc from quasars 
(de Kool et al. 2001; Korista et al. 2008; Borguet et al. 2013; Chamberlain 
et al. 2015). The large size implies a very large mass outflow rate ($10^{2-3}$ 
M$_\sun$~yr$^{-1}$) and large kinetic power (a few to 10 percent of the total luminosity) if 
the outflows cover a substantial fraction of the sky. However, absorption lines 
can only probe the gas on the line of sight, so the global covering factor of 
absorbing gas is virtually unknown in an individual object. Because these 
quasars are so rare,  the average global covering factor must be very 
small, according to the unification scheme, but if they are at a special evolution 
stage, on the other hand, the covering factor can be substantial. 
Interestingly, Liu et al. (2015) detected extended outflow components in 
two low z AGNs on the physical sizes consistent with those inferred 
from ultraviolet absorption lines. 

In this paper, we report four quasars, including one that was noted already by 
Ross et al. (2015), with peculiar ultraviolet \feii spectra.  They were discovered 
during the course of studying a sample of heavily obscured quasars in WISE: 
$W1-W4>8.0$.  
We perform an analysis of the emission lines and the broad band SED, and 
present evidence that the \feii emission lines in these objects are dominated 
by continuum fluorescence pumping in the extended outflows, so they can be 
used to constrain the covering factor and size of BALR of LoFeBAL. 
The paper is organized as follows. In section 2, we present an 
analysis of the optical spectra and broad band SED. In section 3, we discuss 
the nature of \feii emission and continuum emission, and a brief conclusion 
is given in section 4. Spectral models of fluorescent Fe~II emission 
are presented in an appendix.

\section{The data analysis}

\subsection{The Optical Spectra}
\label{2.1}

The optical spectra are shown in Figure 1. One of these quasars, SDSS 
J153542.4+090341 has already been noted in Ross et al. (2015); and another 
one (SDSS J163246.6+340526) appeared on the list of unusual quasars in 
Meusinger et al. (2012). The most striking features of these objects are 
weak continua, strong ultraviolet \feii spikes and \mgii emission lines, 
and a relatively narrow \mgii line. Individual objects are described and 
analyzed as follows.

\begin{itemize}
\item{SDSS J153542.4+090341: 
The BOSS spectrum shows three prominent \feii spikes due to UV1,2,3 
and 62,63, and \mgii, \aliii and \civ lines. \heii 1640+\oiii 1663 
and \ciii] are weak with respective to these lines. We measure an extreme 
\feii spike/gap ratio\footnote{It is defined as Fe II flux over the 
wavelength ranges 2312--2428\AA (hereafter, Fe2B2355)~ and 2565--2665\AA~ dividing by the 
total Fe II flux in the range 2462--2530\AA (Baldwin et al. 2004).} of
8.5.  A narrow \feii 78 is also visible. In fact, the equivalent widths 
of \ciii] and \heii are consistent with those of I Zw 1 (see Figure 1, 
also Laor et al. 1997), while those of \feii, \mgii, \aliii and \civ 
are a factor of 10-20 times  larger. This indicates that the UV 
spectrum consists of two components, a weak I Zw 1 component and 
another component dominated by strong emission lines, \feii, \civ, 
\aliii and \mgii. \civ shows two components, a narrow component at 
the systematic redshift, derived from 
[O {\sc ii}], and a blue-shifted component. Alternatively, \civ may be 
split into two components due to a blue shifted narrow absorption line. 
The \mgii doublet can be well modeled with two gaussians. The best fit 
converges to a width of $FWHM=1230\pm60$\kms and no shift in velocity 
(2$\pm$500\kms). 
[O{\sc II}] appears fairly narrow overall, but with weak broader wings. 
[Ne {\sc iii}]$\lambda$3869 looks similar with a narrow core and extended 
wings. However, both lines are severely contaminated by the subtraction of 
sky lines. }

\item{SDSS J100552.6+493448: 
In addition to strong ultraviolet \feii spikes, this object displays very 
strong high-ionization forbidden and permitted lines: [\oiii]$\lambda$4363, 
[Ne{\sc iii}]$\lambda\lambda$3869,3987, [Ne{\sc v}]$\lambda\lambda$3346,3426, 
[Ne{\sc iv}]$\lambda$2422, [C{\sc iii}]$\lambda$1906,1909, as well as low 
ionization lines $H\gamma$, $H\delta$, [O{\sc ii}]$\lambda\lambda$3726,3729, 
C{\sc ii}]$\lambda$2326, Mg {\sc ii}, and Al {\sc iii}$\lambda\lambda$1855, 
1863. In fact, the spectrum can be reproduced by a supposition of broadened 
SDSS J1535+0903 emission lines and the narrow lines mentioned above (See Figure 1). The 
\aliii strength can be well accounted for by the SDSS J1535+0903 component. 
To  first 
order, all but [O {\sc ii}], have the same width, around FWHM$\simeq$3150\kms, 
using a single gaussian fit. [O {\sc ii}] is considerably narrower than other 
lines, probably from an extended star-forming region. There is also a weak 
broad wing in [O {\sc ii}], which may be similar to other lines. The weak 
line to the red side of \mgii can be either \feii or Mg {\sc i}. 
[\oiii]$\lambda$4363 is stronger than H$\gamma$ according to the single 
gaussian fit with their centers tied to other lines. Unfortunately, the BOSS 
spectrum does not cover [\oiii]$\lambda\lambda$4959,5007, which should impose 
important constraints on the gas density and gas temperature in combination 
with [\oiii]$\lambda$4363. Another constraint on the density comes from 
the fit to 1900 blending. We find that in the single gaussian model for 
each line the fit is significantly worse when the lines are assumed to be 
[Si {\sc iii}] and [\ciii] with a $\Delta\chi^2=30.8$ with respect to models 
assuming they are Si {\sc iii}] and C {\sc iii}]. A further fit with a mixture 
of \ciii] and [\ciii] yields a [\ciii]/\ciii]=0.11$\pm$0.02 with a 
$\Delta\chi^2=11.8$ lower than pure \ciii] model. This gives a gas density 
$(3-4)\times 10^5$~ cm$^{-3}$ (Osterbrock \& Ferland 2006). The outflow has 
a considerably lower density than that in Q 1321+058 (Wang et al. 2009). 
Detailed photoionization modeling of the outflow component will be presented 
elsewhere. }
 
\item{SDSS J163246.6+340526: The spectrum is dominated by narrow ultraviolet 
\feii lines. The \feii spikes/gap ratio (3.2) is less extreme than in 
J1535+0903. 
We adopt the redshift of narrow [O {\sc II}] as the systemic redshift. 
\feii and \mgii appears fairly narrow. We fit the \mgii multiplet using 
two gaussian and obtain a best fit with an $FWHM=880\pm40$\kms and a doublet 
ratio of $I(2796):I(2803)=0.48\pm 0.04$ with no systematic redshift. \mgii~is 
slightly narrower than \mgii in I Zw 1 (Laor et al. 1997). Interestingly, 
the observed spectrum in the 2100-3100\AA~ range can be reproduced 
approximately by the addition of a scaled I Zw 1 spectrum with four-times 
stronger emission 
lines plus the \feii\ spectrum of SDSS J1535+0903 (see Figure 1c). [O {\sc iii}]
$\lambda$5007 is visible in the gap of the sky line forest, and has the same 
redshift as [O {\sc ii}].  Unfortunately, H$\beta$ is seriously affected by 
residuals of sky lines. }

\item{SDSS J212546.9+004455: Unlike other objects, the continuum emission in 
this object is relatively strong. The narrow \mgii line has the same redshift 
as [O {\sc ii}], which is considered as the systematic redshift. A comparison 
with the \feii template of I Zw 1 and the spectrum of SDSS J1535+0903 suggests 
the presence of two blue-shifted ultraviolet \feii components. A broadened 
SDSS J1535+0903-like component is shifted by 5700 \kms~ and a I Zw 1-like 
ultraviolet \feii component is blueshifted by $\sim$2850 \kms. 
The I Zw 1-like component has weaker \mgii, \ciii, Si {\sc iii} and \civ lines 
but broader line widths than I Zw 1. We measure the FWHM of Mg {\sc ii} using 
the above two-component fit (see also Figure 1). There appear narrow absorption 
lines of \civ at velocities of 900 and 2300 \kms. But the signal to noise ratio is low.} 
\end{itemize}

Table 1 presents the emission-line parameters, including the width of \mgii\ 
and the equivalent widths of \feii spikes, \mgii and \aliii. 
To summarize, SDSS J1535+0903 represents a prototype  peculiar \feii 
spectrum, while the spectra in other objects can be modeled as a combination of 
this and a I Zw 1-like \feii component.  
  
\subsection{Modeling the Spectral Energy Distribution}
\label{2.2}

Broad band SEDs of these quasars are shown in Figure 2. They are characterized 
by strong and steep mid-infrared emission, and a relatively weak ultraviolet 
and 
optical continuum. The photometry is strongly affected by the ultraviolet \feii 
and \mgii. We corrected the emission-line flux using the SDSS spectrum. 
We fit these SEDs using a model consisting of a clumpy torus  
plus a galaxy template, accounting for the host galaxy light. We use all 
galaxy templates, excluding AGNs, that are in the SWIRE library (Polletta et al. 2007). 
The purpose of this fit is to give an estimate how much  of the 
infrared emission comes from the AGN heated dust and how this dust is distributed. 

The SEDs of clumpy tori were calculated for a wide range of parameters by 
H\"onig \& Kishimoto (2010). In their models, the torus consists of optically 
thick clouds with a small total filling factor. The number density of clouds 
decreases outward in a power law form, and has a gaussian distribution in the 
vertical direction. The parameters of the models are the number of clouds 
along the line of 
sight on the equator ($N_0$), the radial dust-cloud distribution index 
($\alpha$), the subtending angle of the torus ($\theta_0$), the optical depth 
of  individual clouds, and the inclination (refer to H\"onig \& Kishimoto 2010 
for details). The inner radius is fixed to the dust sublimation radius and the 
outer radius of the torus to 300 dust sublimation radii.

The broad-band SED can be fitted fairly well with such a model. The best 
fits are shown in Figure \ref{sedfit} and the parameters are listed 
in Table \ref{emissionline}. We also compute the distribution of the AGN 
fraction in the rest frame 3-15 $\mu$m using all models weighted 
by the $\chi^2$ probability of each model and by the solid angle of the 
viewing angle, i.e., $w=P(\chi^2,dof)\cos\ i \Delta i $.
The results suggest that the infrared emission is dominated 
by a torus component, and that the galaxy component contributes less than 
5\% in all cases. The power-law index for the dust 
distribution is quite broad from $-2.0\geq \alpha \geq -0.7$. Thus the 
steepness in the mid-infrared spectrum is due to weak host galaxy light 
in the near-infrared, indicating a high Eddington ratio of the system, 
and an edge-on viewing angle. This result is probably a color selection 
effect, where we only choose objects with a very red infrared color. Second, 
the half opening angle of the torus is small (45 to 30 degree), and the 
system is viewed at a relatively large inclination angle (65-75 degree), 
consistent with the large infrared to optical ratio. Finally, torus models 
predict a UV flux higher than the observed ones if there is no internal 
extinction. The reddening to the reflected light can be due to diffuse 
interstellar dust further out in the host galaxy, which may be partially 
covering. Note that our model does not include other optical or UV continuum 
sources, such as direct or electron scattered accretion disk 
light. If such components are present, then the contribution from the host 
galaxy should be even lower and the optical and UV continuum from the 
dusty torus should be highly extinguished. 

\section{Discussion}

\subsection{On the Continuum Emission}
\label{3.1}

The large infrared to optical/UV ratio and the steep near to mid-infrared ratio
indicates that the optical to UV, and even the near-infrared, emission from 
the accretion disk is obscured. Assuming the mid-infrared emission, 
which is dominated by the dusty torus emission as suggested by SED 
fitting, is nearly isotropic, the continuum emission at rest frame 
3000\AA~ is a factor of 30-300 weaker than in normal quasars (refer 
to Figure 1). If the 3000\AA~flux is attributed to the transmitted light, 
the dust extinction would be $E(B-V)=0.65--1.06$ for an SMC-like 
extinction curve. Applying such dust extinction to the average UV 
spectrum of quasars, we will get a slope $\beta=3.1-6.0$ ($f_\lambda
\propto\lambda^{+\beta}$) between 2000 and 3000\AA, %assuming an intrinsic $\beta=-1.7$), 
which is much steeper than the observed one. Adopting an average Galactic 
extinction curve will yield only slightly smaller slope. This indicates 
that the observed UV continuum is  reflected/scattered light. In fact, 
the residual ultraviolet flux can be well accounted for by  reflected 
light from the clumpy dust torus itself as in the SED fitting. Alternatively, 
the continuum can be due to electron scattering, which is  known to be important
in Seyfert 2 galaxies (e.g., Antonucci \& Miller 1985). This suggests 
that the optical depth to the continuum emission region is much larger 
than indicated by the $E(B-V)$ derived above. In either case, the continuum should 
be polarized, and future polarization observation could verify this. 

In the SED fit, we found that the dusty torus has a large subtending 
angle, i.e., covering 71 to 87\% of the sky. The large inclination angle 
suggests that our line of sight to the continuum source is blocked 
by the torus, consistent with weak optical and UV emission. In this 
case, the hot dust in the illuminated face of inner dust torus is 
also blocked, leading to a steep rise from the near to mid-infrared. 
So the dusty torus can naturally explain the obscuration and infrared 
SED. The SED fitting also suggests that the contribution of stellar 
light to the near and mid-infrared is small, which implies a high 
Eddington ratio.  

Once the direct continuum is obscured, the large line equivalent 
width can be understood if the emission-line region is more extended 
than the dusty torus. In a few nearby Seyfert galaxies, the warm few 
hundred K dust which is responsible 
for mid-infrared emission is resolved into a thick structure of a 
few parsecs with a bolometric luminosity around $10^{44}$ erg~s$^{-1}$ 
(Tristam et al. 2007; L{\'o}pez-Gonzaga et al. 2014; Tristram et al. 2014). 
Note that this is not the outer radius of the torus, which might 
extend to several tens of parsecs. Since the size of the torus is 
proportional to $L^{1/2}$, with an infrared luminosity of 
$10^{46-47}$ erg~s$^{-1}$ for our quasars, the true size  will 
be at least several tens of parsecs. In either case, the emission line 
region should be larger than this.  

Before making a detailed discussion the size of the obscurer, let us consider 
the constraint on the size of emission line region first. The emission line 
width (FWHM), as fitted with \mgii, is in the range  900-8000 \kms, which is 
broader than narrow lines in Seyfert galaxies, but is comparable 
with or narrower than the broad lines. Naively, this can be interpreted as 
lines that come from an intermediate region between the traditional BLR and NLR 
(Hu et al. 2008; Li et al. 2015). However, note that SDSS J1005+4934 also 
displays strong forbidden lines of similar width ($\sim$3000 \kms), different 
from traditional 
intermediate broad lines. Thus, we consider it more likely that these emission 
lines come from the inner region of NLR, and that the large width is due to 
outflows (e.g., Komossa et al.  2008; Zhang et al. 2011).
In this picture the traditional BLR is completely obscured. 

While the dusty torus provides the most naturally explanation for 
the obscuration, 
it is worth considering other alternatives. The obscurer can be a single 
dusty cloud in the clumpy torus model or the distant interstellar medium. The 
obscuration of the BLR requires a minimum size  cloud to be $R_{BLR}\simeq 
0.1(\lambda L_{\lambda,5100}/10^{45} {\mathrm erg~s}^{-1})^{0.5}$ 
parsecs (Bentz et al. 2013). Using the infrared luminosity as an 
approximation for the 
bolometric luminosity and a bolometric correction to $L_{bol}/
\lambda L_{\lambda,5100}\sim9$, we estimate  BLR sizes in the range 
of 0.1-0.3 parsecs.  Assuming a Galactic dust to gas ratio ($N_H/E(B-V)
=(2.0-2.5) \times10^{-22}$ cm$^{2}$, Bohlin et al. 1978; Cox et al. 2006), 
the gas column density of the obscurer should be much greater than 
$4\times10^{21}$ cm$^{-2}$ as discussed above. Note that in the clumpy 
torus, the gas density should be of the order of a few $10^6$ cm$^{-3}$ (Krolik 
\& Begelman 1988), so a BLR-sized cloud would have a column density two 
orders of magnitude larger than the above minimum value, or  
$E(B-V)\sim 100$ Mag for a Galactic dust to gas ratio. On the 
other hand, assuming a gas density more typical of the interstellar medium 
$n_H\sim 10^2$ cm$^{-3}$, the minimum size of cloud would be several 
tens of parsecs to meet the minimum column density. 

In the case of the single-cloud obscurer, one would expect to see long-term 
very large amplitude variability as the cloud moves in and out of the line 
of sight. The typical time scale would be the size of the cloud divided 
by the transverse speed, which is of order the local Keplerian velocity. In 
a typical galaxy, the orbit velocity is order of 500-1000 \kms, depending on 
the distance to the center, and a minimum size of a cloud should be  BLR sized, 
of order of 0.1 to 1 parsecs. This would give a variability time scale 
$\sim 10^{2-3}$ yr. So it is difficult to observe the cloud transit event 
fully. However, the 
continuum source is much smaller (a hundred to thousand times smaller), 
so it is possible to observe transient obscuration of the continuum source 
with small changes in broad lines on time scales of years if the cloud has 
a sharp edge. There can also be cases where the cloud blocks the 
continuum source fully but only partially blocks the broad line region as 
the obscuring cloud moves in and out. These sources will show very 
large equivalent widths of broad lines. The probability of detecting such a
source depends on the relative size of the obscuring cloud relative to the BLR size. 
In the case of comparable sizes of the two, one would expect that in most 
sources only the continuum is fully covered, while if the cloud is much larger 
than the BLR, most likely both continuum and BLR are obscured. Similarly, we will 
observe many more such partially obscured sources in the clumpy torus than 
in the smooth dust torus. In reality, the cloud may have a smooth edge 
so the contrast between emission line and continuum should be much moderate. 
We note that a single cloud occultation event is likely to give 
a smaller polarization degree than the dusty torus obscuration due to 
different symmetries (Marin et al. 2015).

The weakness of the optical to UV continuum and the strong mid-infrared 
emission may be due to a temporarily shut-off of the central engine. A 
comparison of infrared,  optical, and UV fluxes requires a drop in the UV flux 
from the inner disk by a factor of 30-300 in these sources, compared to 
typical AGN (see Figure 2). The size of the mid-infrared emission region 
is of order of tens to hundreds of dust sublimation radii at the infrared 
luminosity according to our model of the SED (H{\'o}nig et al. 2010). At 
the infrared luminosity of these sources, the light-crossing time of the 
mid-infrared emission region is of order  ten to a hundred years, so the 
central engine would have to have shutdown in less than this time. A 
comparison of SDSS photometric data with Palomar Observatory Sky Survey 
(POSS) on a time scale of 50 years in the observer 
frame shows a small fraction of quasars with a variability amplitude of 3 
mags in $g$-band (de Vries et al. 2005). Although these amplitudes are still 
smaller than that is required, since we do not have  light curves of AGN on 
such long time scales, it is impossible to tell whether such a large amplitude 
variability happened on time scale of a hundred years 
and how frequent they are. Thus this remains a possibility. 

There are several tests that could distinguish among these possibilities. 
First, scattered light should be highly polarized, while direct light 
from a much less active accretion disk is likely not.  
The observed scattered light in Seyfert 2 galaxies is polarized by a few 
to tens' percent, while the polarization of direct emission from quasars 
is usually less than 1\% (Marin et al. 2015). It should be noted that if 
a quasar resides in the gas-rich environment, there is inevitable scattered 
light from surrounding gas. As the central continuum source dims quickly, 
the fraction of scattered light increases, and so does the polarization degree. 
So a more critical test would be to observe the transmitted hard X-rays. 
Second, direct emission from accretion disk is likely to be variable on relatively 
short time scales, while the scattered light is not. Thus, the detection of 
variability on time scales of less than a year would support the intrinsic 
weak continuum emission model. 

\subsection{The Origin of Strong \feii Emission}

We now turn to the properties of the emission region. Previous studies 
suggested that quasar emission lines consist of distinct kinematic 
components: broad, intermediate and narrow components (Brotherton et 
al. 1994; Marziani et al. 1996). Hu et al. (2008) found that optical 
\feii lines are formed in the intermediate region, that is redshifted 
with respective to the system. 

The \feii spectrum of our sources does not look like those in normal 
quasars or from most theoretical simulations aimed at reproducing observed 
\feii in I Zw 1 (Baldwin et al. 2004; Sigut \& Pradhan 2003; Ferland et al. 
2009; Bruhweiler \& Verner 2008). We will mainly focus on the Fe {\sc ii} 
emission in SDSS J1535+0903 first, because in other sources the spectra 
can be reproduced with a combination of SDSS J1535+0903 and I Zw 1-like 
components. The UV \feii emission is dominated by a few bumps, namely 
multiplets UV 1, UV 2,3, UV 62,63, with weak emission in between and 
very weak or no emission at wavelengths longer than 2900\AA. In SDSS 
J1005+4934, the SDSS spectrum extends to 4400\AA~in the quasar rest frame, 
and optical \feii lines appear relatively weak. 

For thermal excitation, the extreme spike/gap ratio of 8.5 in SDSS 
J1535+0903 may indicate very high densities ($n_H>10^{13}$ cm$^{-3}$) 
and very low ionization parameters ($U=10^{-4--5}$), as shown in Figure 3 of 
Baldwin et al. (2004). But their models are only for the broad line region, 
and the ionizing photon flux is too large for our case. However, in 
their figure, the curves of iso-spike/gap ratio run diagonally, so the  
dimensionless ionization parameter may be more crucial than the density. 
Unfortunately their models do not extend to the low density and very low 
ionization regime, so it is not clear whether extreme spike/gap ratios 
can be reproduced there. But if this is indeed the case, the \feii 
emission region must be far from the quasar so that it is not 
obscured by the dusty torus. At very low values of the ionization parameter, 
some other lines, such as  C~{\sc ii} and Ni~{\sc ii}, might become strong. 
Detailed photo-ionization modeling would be required. Also the 
presence of strong \civ and \aliii, and the lack of \ciii~and Si {\sc iii} 
cannot be explained by this model. 

Alternatively, the extended emission may come from resonant scattering 
in low-ionization quasar outflows. Previous observations suggest that 
FeLoBALs and HiBALs, at least in some quasars, are formed on the scale 
of 100 pc to a few kpc (see references in \S 1), so it is likely that 
outflows are not obscured by the putative dusty torus that blocks the 
line of sight to the central continuum in Type 2 quasars. The outflows may 
produce \feii via collisional excitation, as well as via continuum 
fluorescence. The contribution of fluorescence relative to  
collisional excitation depends on the optical depth of \feii as well as    
the density and temperature of the gas. In the case of FeLoBAL 
outflows, the density is usually low (order of $10^{3-5}$ cm$^{-3}$, 
de Kool et al. 2001; Lucy et al. 2014; c.f., Zhang et al. 2015), and 
the optical depth of most \feii lines is usually not large (order of 
a few to tens for strong lines). Thus, continuum fluorescence may be the 
dominant process in the outflows for \feii lines. 

\begin{figure}
\centering{
 \includegraphics[angle=90,origin=c,scale=0.50]{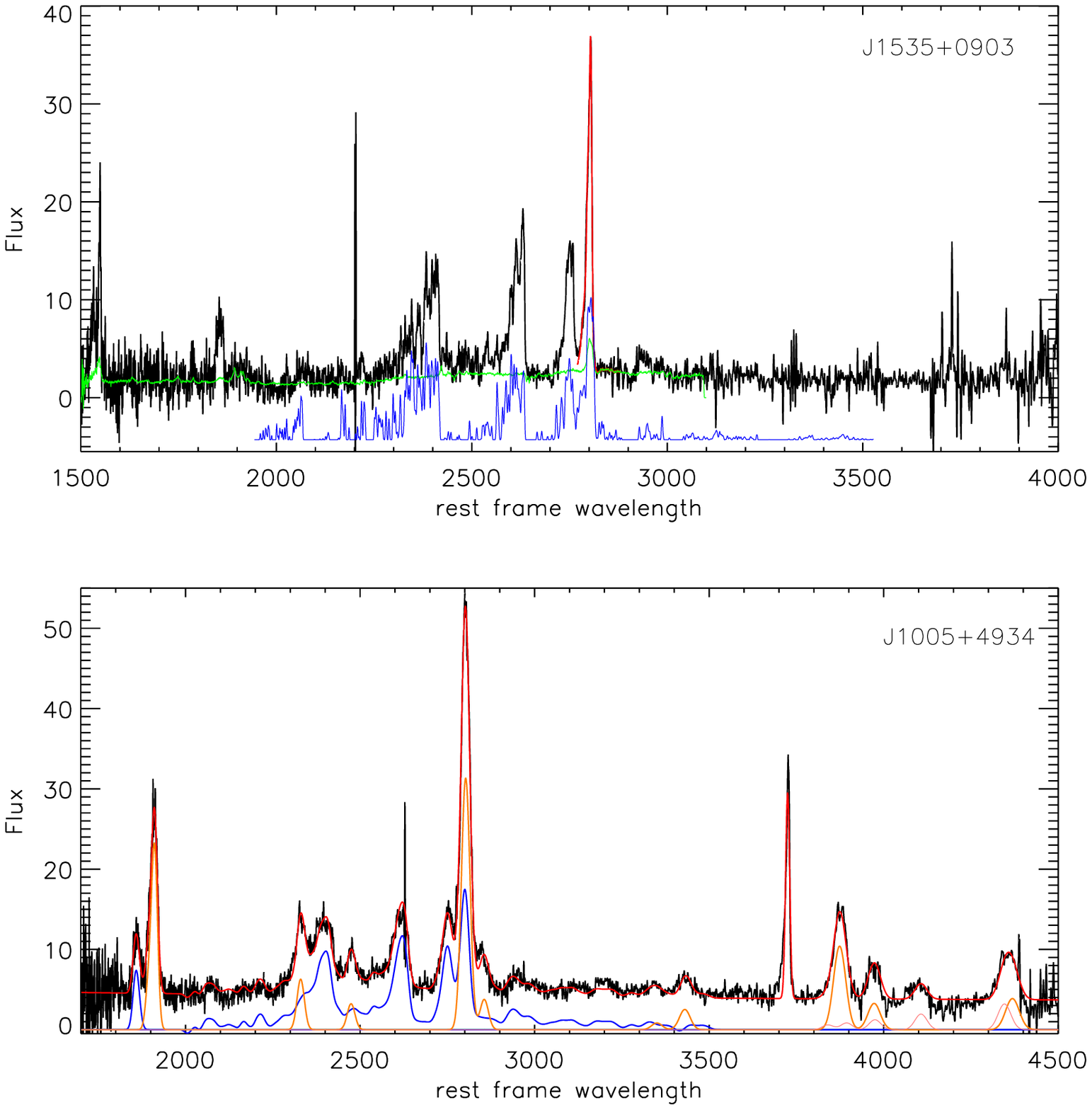}
 \includegraphics[angle=90,origin=c,scale=0.50]{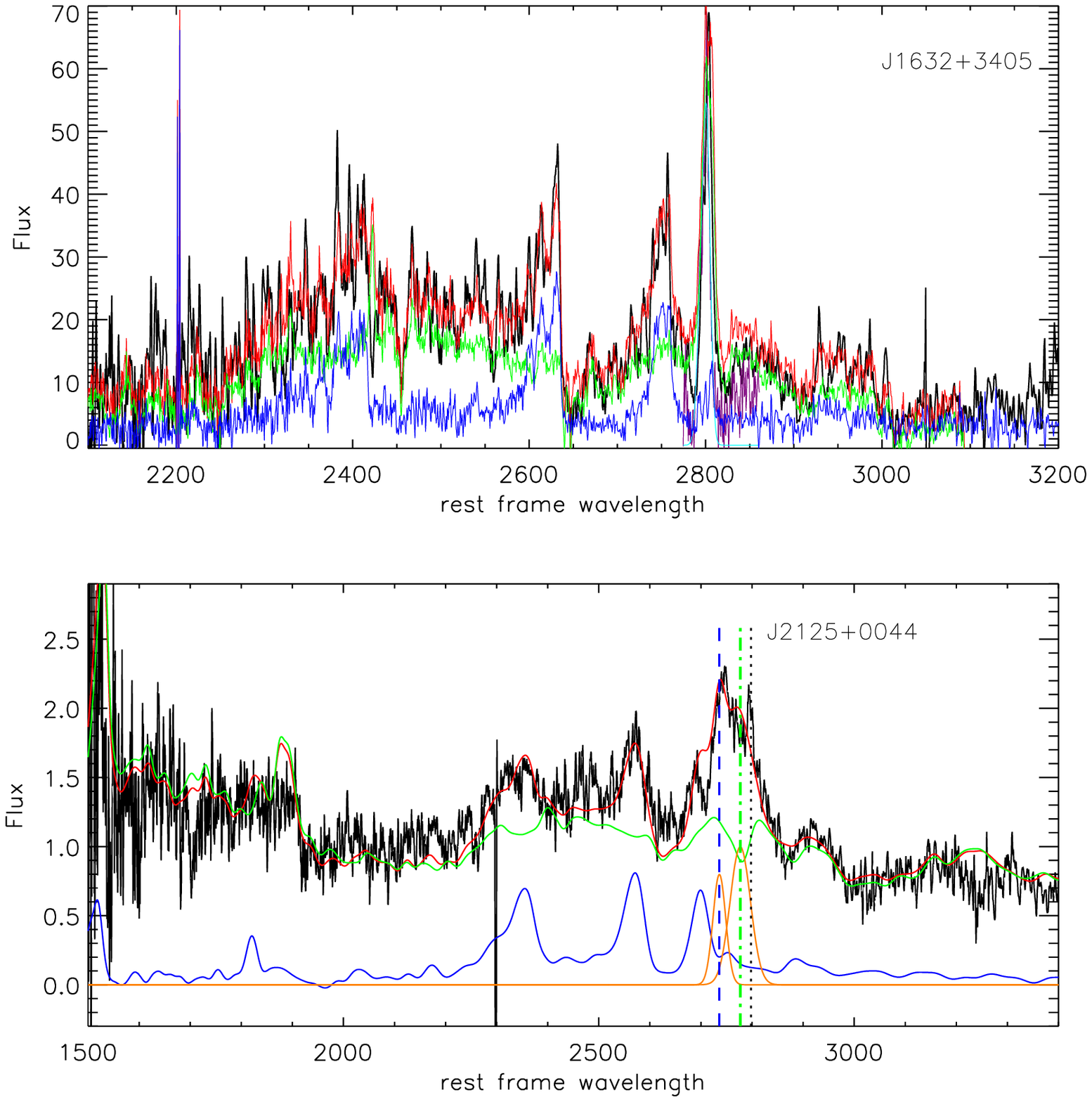}
 \caption{The SDSS spectra of four AGN with peculiar Fe {\sc ii}  emission (in black 
curves). In the top panel, the blue curve is the absorbed spectrum of FeLoBAL 
QSO SDSS J080957.39+181804.42, which is derived using the best matched 
composite templates (See Wang et al. 2015 for details). The \feii spectrum 
of SDSS J1535+0903~looks remarkably similar to the absorption spectrum of 
FeLoBAL QSO. In the second panel, the blue curve is the broadened SDSS 
J1535+0903 spectrum, while orange 
curves are gaussians for additional emission lines. In the third panel, the green 
curve represents the scaled I Zw 1 spectrum, the blue is the scaled SDSS J1535+0903~ 
spectrum, and the red one is the combination of the two. In the bottom panel, colored
curves are as in the third panel except that the \mgii line in each component 
is replaced by an independent gaussian. The SDSS J1535+0903 component is 
blue-shifted by 6600\kms, and the I Zw 1 component by 2300\kms.
\label{specfit}}
}
\end{figure}

\begin{figure}
\centering{
 \includegraphics[scale=0.90]{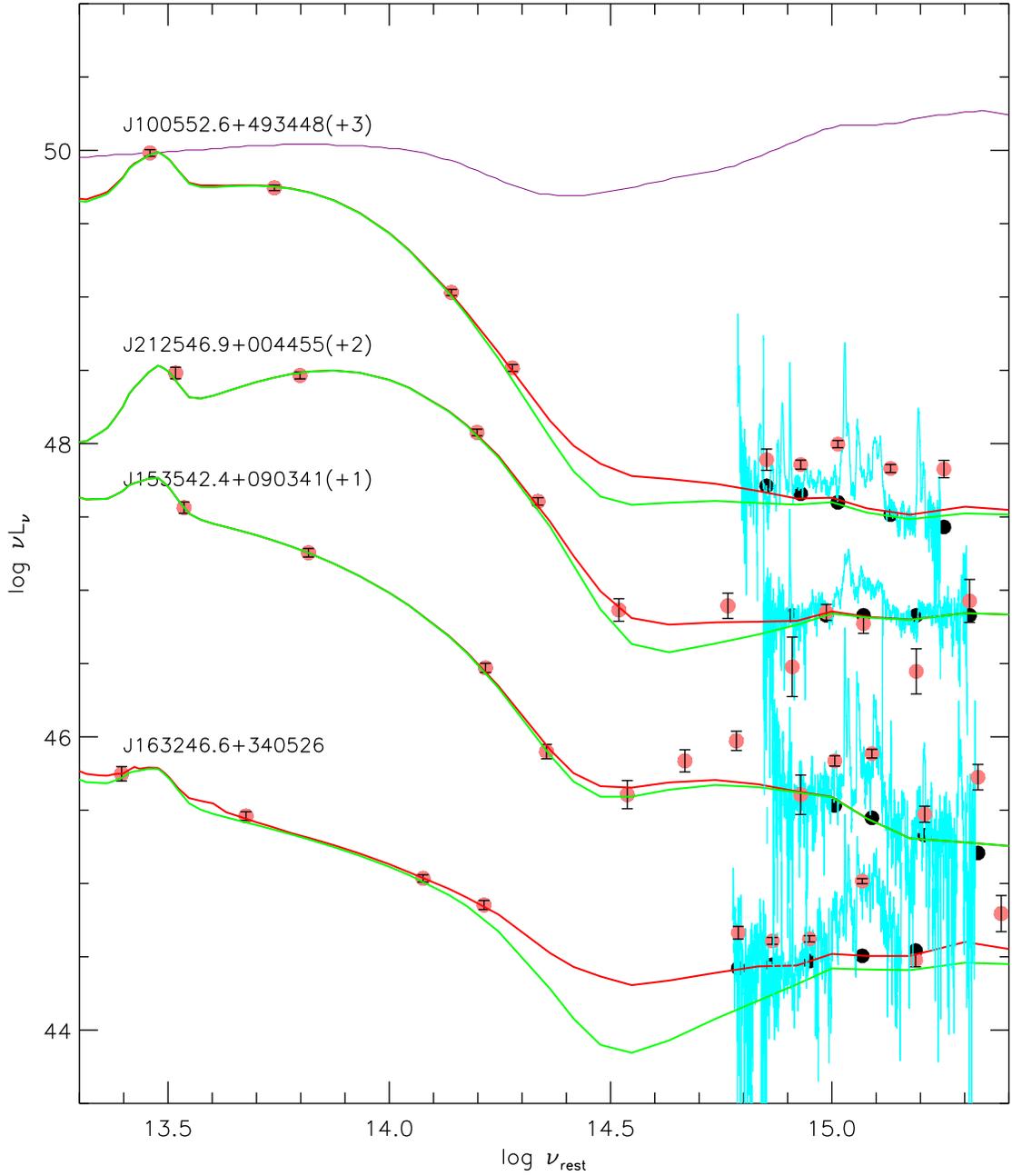}
 \caption{The spectral energy distribution and best-fitted torus model 
for four quasars are shown. 
The filled red circles are photometric data from WISE, UKDISS and 
SDSS, the black ones are those matching the SDSS spectrum 
(shown in cyan). The red curve is a best fit with a combination of a 
reddened dusty torus and a galaxy template; the green curve is the 
reddened torus component. For comparison, we also show the composite 
SED for red quasars from Richards et al. (2006) in purple.
\label{sedfit}}
}
\end{figure}

Some other emission lines may be also dominated by 
resonant scattering from such outflows. Consequently, these lines should appear strong in 
comparison with non-resonant lines. The presence of strong \aliii, 
but very weak or missing similar component of \ciii] and Si {\sc iii}], 
which are usually much stronger, in SDSS J1535+0903, indicates that 
resonant scattering is the dominant excitation mechanism in this 
object. \aliii can be produced by either resonant scattering or collisional 
excitation, while \ciii] and Si {\sc iii} are only collisionally excited. 
Although the relative strength of \aliii increases with gas metallicity, 
this does not explain the fact that \aliii is much stronger than 
\ciii]. In Table 2, the \aliii equivalent width is a factor of 3.7 larger than the 
\ciii] equivalent width. As discussed in \S \ref{3.1}, the observed UV 
continuum is likely scattered light, and, as seen in Figure \ref{specfit}, 
the equivalent widths of \ciii] and Si {\sc iii} is similar to that of I Zw 1. 
So \ciii] emission may be fully accounted for by scattered component and 
the contribution from the outflow component may be negligible.

If \feii is produced by continuum fluorescence in outflows, a simple idea 
is that its spectrum may resemble those of the \feii absorption line spectra 
in FeLoBAL QSO SDSS J080957.39+181804.42. We note that there are usually 
very weak \feii absorption lines above 2800\AA, and no optical \feii 
absorption lines, consistent with the observed \feii in these spectra. 
To test this idea, we extract the absorption line spectrum of a FeLoBAL 
QSO. This spectrum is shown in Figure 1, and is similar, 
to  first order. It should be noted that the spectrum of 
fluorescent emission is not exactly the same as the \feii absorption 
spectrum, because 
an \feii ion in an excited level, after absorbing a photon, may decay 
to a different lower level, thus emitting a photon with an energy different 
from the absorbed one, especially when resonance lines become optically 
thick. Detailed calculations are presented in \S \ref{model}.  

In Appendix A, we demonstrate that a pure fluorescent model can indeed  
reproduce the observed \feii band ratios in SDSS J1535+0903 very well. 
In order to make \feii\ the dominant ion of Fe, we let the incident 
ionizing continuum be filtered by an intervening absorbing gas of column 
density 10$^{23}$~ cm$^{-2}$. We adopt a gas density $n_H=10^{4.5}$ cm$^{-3}$, 
consistent with those in the FeLoBAL QSOs, and a turbulent velocity of 
1000 km~s$^{-1}$ to mimic the outflow velocity. We find that at the optical 
depth of the resonant \feii line at $\tau(2599A)$ is 900, and that 
the model produces UV FeII band ratios that are consistent with the observed 
ones. As far as we know, this is the only model that gives such an extreme 
\feii ratio at a low ionizing continuum flux, which is consistent with its 
large line formation region\footnote{Some models in Baldwin et al. (2004) 
also show such extreme line ratios but at high ionizing photon flux.}. 
Furthermore, the model gives weak \feii lines at wavelength longer than 
2800\AA, which is also consistent with the observation. 
The best matched model predicts an equivalent width of Fe {\sc ii} 
at 2355\AA~band of 99\AA, which is about one 6th of the observed value 
for SDSS J1535+0903, for a covering factor of one. However, as we discussed 
in \S \ref{3.1}, the continuum incident to the gas is obscured and the observed 
continuum is attenuated by two orders of magnitude, so only a few percentages 
covering factor would be sufficiently to explain the observed Fe {\sc ii} 
strength. This covering factor is consistent with the estimate in next 
subsection. The gas producing the Fe II emission is cool, as shown in the 
appendix, so collisional excitation is unimportant.  This accounts for the
weak optical Fe {\sc ii} emission. 

Note that this type of model will also predict a number of other low 
ionization lines, such as Ti {\sc ii}, Ni {\sc ii} and Si {\sc ii}, 
but their atomic data are very sparse. So a full treatment of these 
lines has to wait for a major update in the atomic data.
Our models in Appendix A do not produce Al {\sc iii} and C {\sc iv}, 
because ionizing photons above 13.6 eV are required to produce these 
ions while in our scenario these photons are intentionally filtered out 
to prevent destruction of Fe$^+$ ions. Both \aliii and \civ are observed 
in absorption in a FeLoBAL QSO, and their average emission spectrum 
should not be different from the absorption line spectrum in a FeLoBAL 
QSO because they are scattered resonantly. So our purpose is only to 
demonstrate the effect of \feii radiation transfer after \feii resonant 
absorption.  

Note also that the observed \feii ratios from different \feii bands depends 
on the 
reddening, because FeloBAL QSOs often show a large internal dust 
extinction. We do not apply such a reddening correction because 
we do not know the location of the dust. If dust lies outside of the 
BALR, then an
extinction correction should be applied. On the other hand, if dust is within 
the BALR, no correction is needed. Anyway, to  
first order the absorbed line spectrum of FeLoBAL quasars are similar 
to the quasars observed in the sample. 

In SDSS J1005+4934, we also detect strong high ionization forbidden or 
semi-forbidden lines that are excited by collisional process, while the  
\feii spectrum as well as \feii/\aliii ratio, are nearly the same 
as SDSS J1535+0903. We also obtained an electron density of (3-4)$\times 
10^5$ cm$^{-3}$ from \ciii~and Si {\sc iii}. These lines are rather 
broad with widths of 3000 \kms. The presence of such lines does not 
necessarily contradict our interpretation of the strong \feii, \aliii 
and \mgii lines as resonantly  
scattering from an outflow. According the the recent model of 
Faucher-Gigu{\`e}re et al. (2012), 
the FeLoBAL may be produced in the shocked cooling gas during interaction 
of outflows with the cold gas. The gas producing strong high ionization 
forbidden lines may be the warm phase of such shocked gas, and the large width 
of the line may be explain in this way as well. 

Blue shifted \feii and \mgii lines are seen in the SDSS J1632+3405, while 
emission lines in other quasars do not display systematic blueshift. 
The blue shifted lines can be easily explained as from an outflow since 
one expects that the far side of the outflow is subject to large dust 
extinction and would not be seen. However, emission lines from outflows are not necessarily 
systematically blueshifted if the farside of the outflows is not particularly 
obscured or the system is seen nearly edge-on. In the continuum fit, 
we found that SDSS J1632+3405 has the smallest inclination angle among 
these objects. 

\subsection{Constraints on the Covering Factor for the Scattered Region}

If the \feii in these objects is due to fluorescent excitation, we can 
constrain the global covering factor of the outflows in these objects. 
To  first order, we can assume an isotropic angular redistribution 
function, so the scattered \feii flux will be
\begin{equation}
f_{sc}(\lambda)=[1.-exp(-\tau_\lambda)]f_c(\lambda)*C
\end{equation}
where $f_c$ is the un-absorbed continuum flux. Assuming a maximum 
$\tau>1$ at certain $\lambda$,  one can then estimate a minimum covering 
factor $C\simeq [f_{sc}/f_c(\lambda)]_{max}$. To estimate $f_c$, we 
assume that the intrinsic SED is similar to the mean quasar SED, and 
scale the latter to match the WISE luminosity at 22$\mu$m, the peak 
wavelength of SED. By doing so, we obtain covering factors of 0.05, 0.20, 
0.01, and 0.06 for SDSS J2125+0044, SDSS J1632+3405, SDSS J1005+4934 and
J1535+0903. Note that if Mg {\sc ii} is dominated by  resonant 
scattering in the BALR, then the covering factor of Mg {\sc ii} BALR 
would be a factor of 2-3 times larger than the above value. There are several 
uncertainties in this estimate. 

First, due to large number of energy levels, Fe {\sc II} cannot be rigidly 
considered as resonant scattering, but rather as the fluorescent process. 
The absorbed photons may emerge at other wavelength, so the strong resonant 
lines are usually weaker than in the case of resonant scattering. 
Second, it is possible that infrared emission of these quasars is enhanced 
because we selected  objects with very red infrared colors. In this case, 
the unabsorbed continuum is over-estimated. Since SED modeling suggested 
that infrared emission is dominated by the AGN component, while it was shown 
that, on average, thermal dust emission accounts for 35 percent of the total 
power of quasars (e.g., Ma \& Wang 2013), thus at most, the intrinsic continuum 
is over-estimated by a factor of 2-3. Third, the scattered photons may be 
destroyed by dust within or outside the BALR, so we may underestimate the 
covering factor. It has been known for quite a long time that FeLoBAL QSOs 
are reddened. But the location of the dust is still unknown. Recently, there 
are arguments that dust may be mixed within the BALR because near-infrared 
emission is correlated with the outflow velocity (Zhang et al. 2014), and 
the variability of BAL is correlated with the UV spectral slope (He et al. 2015). 
Further, Dunn et al. (2015) showed that reddening must be caused by large 
scale dust because both the continuum and narrow lines are reddened. If dust is 
outside of the BALR and the scattered light  suffers  the same extinction 
as the continuum source, and adopting a median value for the continuum extinction 
for LoBAL $E(B-V)=0.05$ (Dunn et al. 2015), the correction will raise the 
covering factor by about 40\%. Despite these uncertainties, we note that 
in 3 of 4 objects, the uncorrected covering factors are already larger than 
0.05, while according to statistics, FeLoBAL QSOs account for no more than 2\% 
of total quasars (Trump et al. 2006; Dai et al. 2012). This suggests strongly 
that FeLoBAL QSOs must be in a special phase of quasar evolution, rather 
than representing a very small covering factor. 

\section{Conclusions}

We presented an analysis of four quasars with peculiar UV \feii spectra.  They had
extremely strong UV1,3,4,62,63 and very weak \feii at wavelengths longer 
than 2800. Their SEDs rise very steeply from the optical to the mid-infrared, 
suggesting they are obscured. The obscurer is most likely the dusty torus, 
since that also provides a self-consistent explanation for the infrared emission 
although other alternatives cannot be ruled out. The characteristics of 
the \feii spectra can be understood as due to resonant scattering in an 
extended outflow since the size of the scattering region must be outside 
the obscuring torus  and its \feii spectrum similar to absorption spectrum 
of FeLoBALs. Our fluorescent models can successfully produce the observed UV 
\feii band ratios and the weakness of \feii lines at wavelengths longer than 
2800\AA with a low ionizing photon fluxes. This is because the strong UV
lines are excited by continuum pumping within gas that is too cool for
collisional excitation to be significant.  Strong \aliii and weak \ciii] 
and Si {\sc iii} emission may be explained as resonant scattering, as is 
observed in FeLoBAL 
QSOs. The obscuration of the direct continuum also explains the very large 
equivalent widths of \feii, \mgii and \aliii lines. These objects 
are misaligned FeLoBALs. By considering the amount of scattered light we set a lower 
limit on the covering factor of FeLoBALs to be 5\% to 20\% in three 
objects, supporting the scenarios that FeLoBAL QSOs are a special stage 
of quasar evolution.

\begin{acknowledgements}
We thank the referee for constructive comments that help to improve the 
presentation of the paper.
We acknowledge the financial support by the Strategic Priority Research
Program "The Emergence of Cosmological Structures" of the Chinese Academy 
of Sciences (XDB09000000), NSFC (NSFC-11233002, NSFC-11421303, U1431229) 
and National Basic Research Program of China (grant No. 2015CB857005).  
Funding for SDSS-III has been provided by the Alfred P. Sloan Foundation, 
the Participating Institutions, the National 
Science Foundation, and the U.S. Department of Energy Office of Science. The 
SDSS-III web site is http://www.sdss3.org/. SDSS-III is managed by the 
Astrophysical Research Consortium for the Participating Institutions of the 
SDSS-III Collaboration.
GJF acknowledges support by NSF (1108928, 1109061, and 1412155), NASA (10-ATP10-0053, 
10-ADAP10-0073, NNX12AH73G, and ATP13-0153), and 
STScI (HST-AR- 13245, GO-12560, HST-GO-12309, GO-13310.002-A, 
HST-AR-13914, and HST-AR-14286.001).
\end{acknowledgements}

\begin{deluxetable}{ccccccccccccccc}
\tabletypesize{\scriptsize}
\tablewidth{0pt}
\tablecaption{Emission Line Measurements \label{emissionline}}
\tablehead{ 
\colhead{SDSS J} & \colhead{z} & \colhead{FWHM(Mg {\sc ii})} & \colhead{W(Fe2B2355)} & \colhead{$R_{spike}$}  
& \colhead{W(Mg {\sc ii})} & \colhead{W(Al {\sc iii})} & \colhead{W(C {\sc iii})} &  \colhead{W(C {\sc iv})} \\
\colhead{} & \colhead{} &\colhead{km~s$^{-1}$} &\colhead{\AA} &\colhead{} &\colhead{\AA} &
\colhead{\AA} & \colhead{\AA} &\colhead{\AA} }

\startdata
100552.6+493448 & 1.1217 & 3150 & 173 & 6.4 & 310  & 35  & 103 & \nodata \\
153542.4+090341 & 1.5330 & 1230 & 545 & 8.5 & 782  & 186 & 49 &330 \\
163246.6+340526 & 0.8306 &  880 & 280 & 3.2 & 140  & \nodata & \nodata & \nodata \\ 
212546.9+004455 & 1.4240 & 8300 &  82 & 2.9 & 103  &  3 & 6 & 2 \\
\enddata
\tablenotemark{a}{Typical errorbars for strong lines are 20\% for equivalent 
widths and 15\% for FWHM, except for J2125+0044, where line width and equivalent 
width of MgII are poorly determined due to blending with Fe II lines.}
\end{deluxetable}
\begin{deluxetable}{ccccccc}
\tabletypesize{\scriptsize}
\tablewidth{0pt}
\tablecaption{Torus Models for Spectral Energy Distribution\label{sedmodel}}
\tablehead{
{SDSS J} & \colhead{$f_{gal}$} &\colhead{inclination}  
& \colhead{$\Theta_0$} &\colhead{$\alpha$} &\colhead{$\tau_{cl}$} \\
&  & \colhead{deg} &\colhead{deg} &\colhead{} &\colhead{} }
\startdata
100552.6+493448&0.048 & 15 & 60 & -1.5 & 30 \\
153542.4+090341& 0.003 & 15 & 45 & -0.5 & 80 \\ 
163246.6+340526 & 0.086 & 35 & 60 & -1.0 & 80 \\
212546.9+004455 & 0.005 & 15 & 60 & -2 & 50 \\
\enddata
\tablenotemark{a}{Columns are (2) fraction of galaxy light in 2-10$\mu$m band, 
(3) the angle between the line of sight and the equator of torus; (4) the 
half subtending angle of torus; (5) the power-law index of cloud 
distribution as a function of distance to the AGN; (6) the optical depth 
of individual cloud in $V$.}
\end{deluxetable}

\appheading

\appendix
\subsection{Fluorescent Fe {\sc ii} emission}
\label{model}

Here we examine the properties of a cold cloud where iron is mainly singly 
ionized.  Fe {\sc ii} emission will be produced by continuum fluorescence under 
these circumstances.  Our scenario is that an intervening cloud has 
extinguished the AGN continuum and that this attenuated SED strikes an 
outer dust-free cloud.  The intervening cloud is assumed to have a column 
density of N(H) = 10$^{23}$ cm$^{-2}$ and an unattenuated ionization 
parameter of log $U = -1.5$ (although this has little affect on our 
predictions).  The SED transmitted through this cloud, which
then strikes the outer cloud, is shown in Fig A1.

We assume that the outer cloud is dust free. This is to allow Fe to exist 
in the gas phase. Fe is strongly depleted in dusty gas (Jenkins 2009) so 
little Fe {\sc ii} emission occurs when grains are present.  It is possible that 
the cloud was at one time too close to the AGN for dust to exist.  

The cloud would be similar to a classical PDR (Rollig et al. 2007) 
were grains present.  In a PDR grains heat the gas while absorbing the 
Balmer continuum of the incident SED.  Since grains are, by hypothesis, 
absent in our cloud, the gas will be heated mainly by photoelectric absorption by the 
gas, mainly of second and third row elements with ionization potentials smaller 
than hydrogen.  A low gas kinetic temperature results.  The radiation 
striking the gas is bright at wavelengths longward of the Lyman limit, so 
the main species that will be present in the cloud will be H I, He I, 
C II, N I, O I, Ne I, Na II, Mg II Al II, Si II, S II, and Fe II.  Typical 
gas temperatures in our resulting models are around 300 K (Figure \ref{figA3}), too cool to 
collisionally excite optical or UV lines.  UV line emission is produced by continuum 
fluorescent excitation in this case.  Note that it is quite likely that 
the SED striking the cloud is not visible from our line of sight.  This 
would occur if parts of the torus lie between us and the central object.

We use version C13.03 of Cloudy, a spectral simulation code last described 
by Ferland et al. (2013).  Verner et al. (1999) describe the Fe II model 
in this version of Cloudy.  Fe II emission is complex (Baldwin et al. 2004) 
and we report it using the sums over wavelength bands described in that 
paper.

Results are mainly sensitive to the column density of the outer cloud and 
we will show predictions resulting from varying this parameter below.  In 
this cool environment, Fe II is excited mainly by absorption of the incident 
SED by UV lines whose lower levels are close to, or within, the ground term. 
The level populations within the ground term are affected by the density 
which then affects which UV lines are efficiently pumped. The density does not 
affect the emission from high levels so long as it is below the critical 
density of the excited levels, $\sim 10^{10} - 10^{14}$ cm$^{-3}$.  At lower 
densities the emission spectrum is mainly determined by the branching 
ratios as electrons decay to lower levels.  We experimented with a range 
of densities but settled on $\log N(H) = 4.5$ cm$^{-3}$.

\begin{figure}
\centering
\plotone{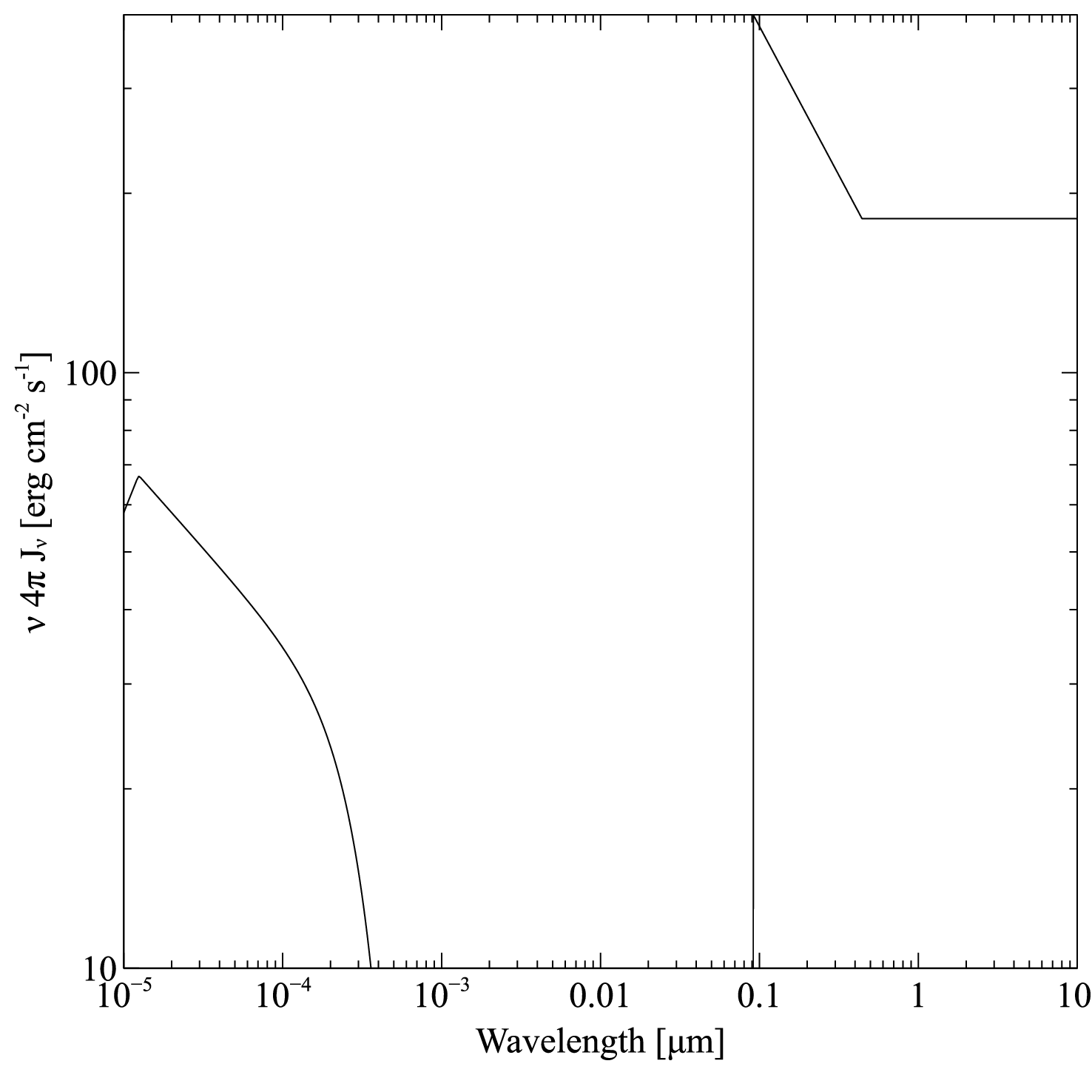}
\caption{The SED striking the outer cloud.  It is a simple AGN SED 
extinguished by a column density $N(H) = 10^{23}$ cm$^{-2}$, which removes 
light shortward of the Lyman limit.  The conditions in the cloud 
are mainly affected by the Balmer continuum.\label{figA1}}
\end{figure}

\begin{figure}
\centering
\plotone{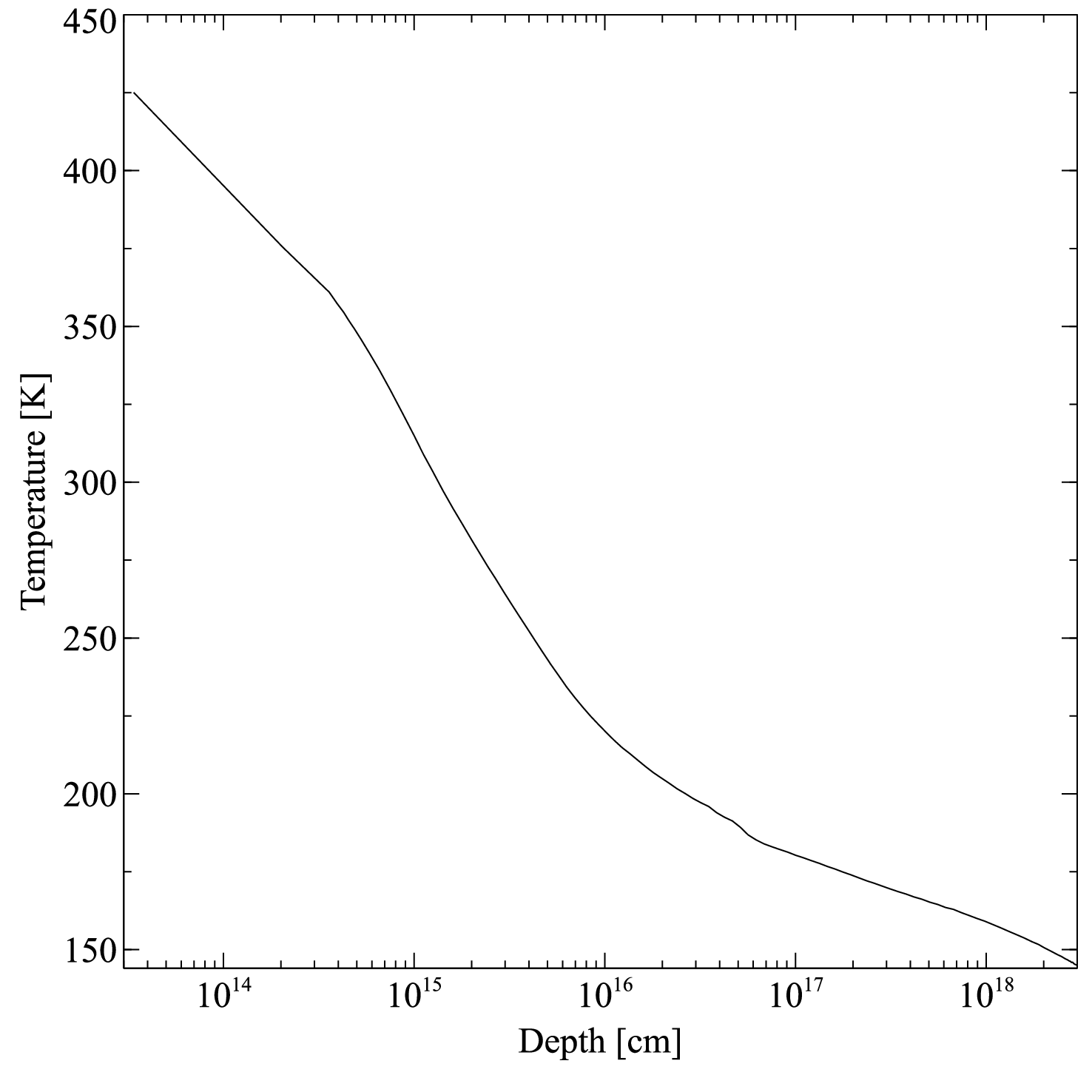}
\caption{The computed temperature structure.  The vertical 
axis gives the temperature [K] while the horizontal axis gives the 
depth into the cloud in cm.  The illuminated face of the cloud reaches 
temperatures a bit over 400 K while the shielded face is much cooler, 
approaching 100 K.  These temperatures are all too cool to permit 
collisional excitation of optical or UV Fe II lines.\label{figA3}}
\end{figure}

Lines absorb mainly over their Doppler core so the amount of pumping will 
depend on the line width (Ferland 1992).  We assume 1000 \kms in the
examples we show although this does not greatly affect the details. The 
Fe II grows stronger as the column density increases and Fe II lines 
absorb more of the incident continuum.  Optical depths in the strong UV 
lines do affect the branching ratios.  As they grow more opaque decays 
from their upper levels will favor subordinate lines.  This is quite 
analogous to the transition from Case A to Case B in H I emission 
(Osterbrock \& Ferland 2006). Finally the intensity of the radiation 
field give in Figure \ref{figA1} must be set.  We assume an unabsorbed 
ionization parameter of $\log U = -1.7$.

\begin{figure}
\centering
\plotone{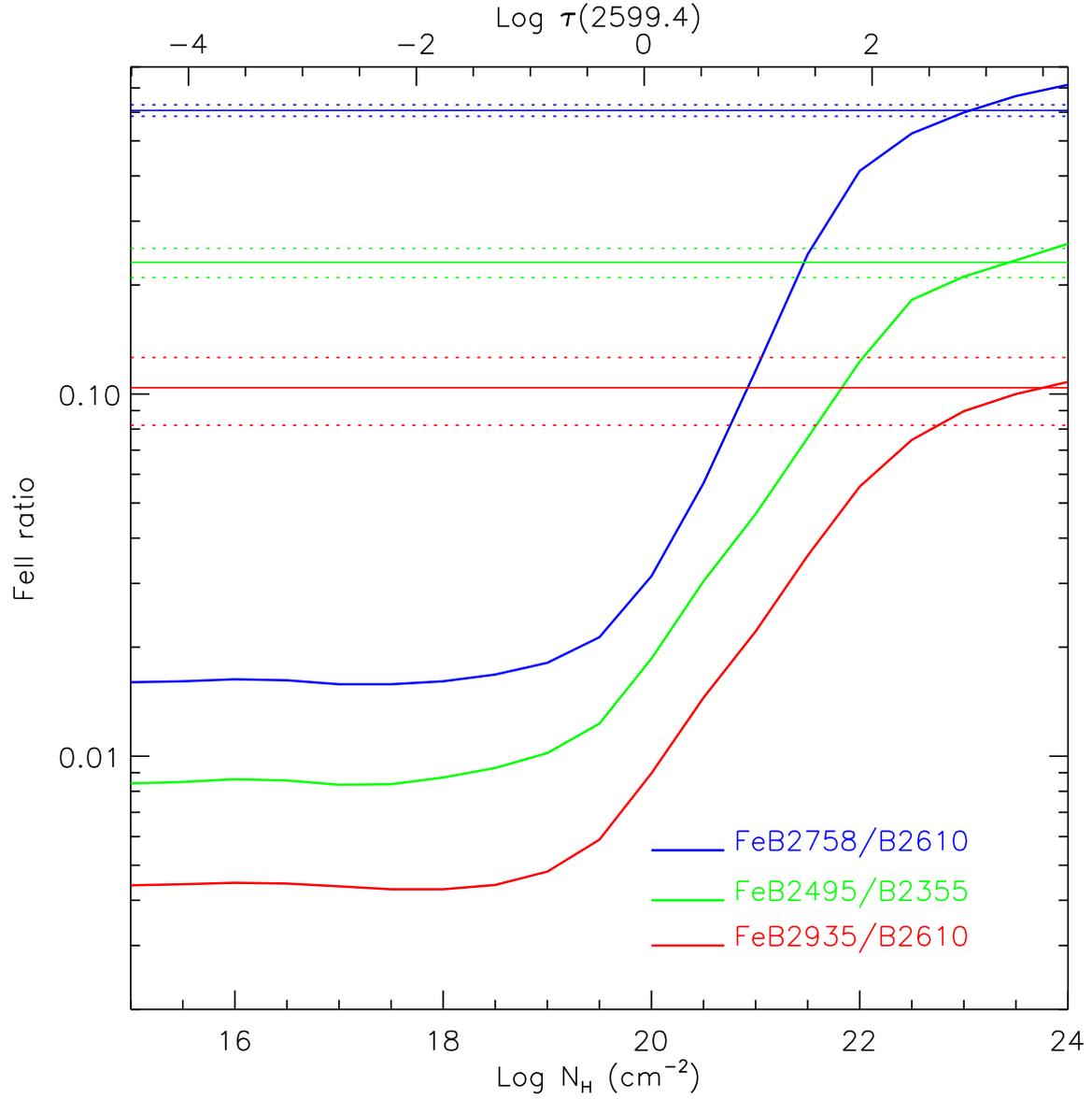}
\caption{Band ratios as a function of total hydrogen column density 
on the bottom and the optical depth in a strongly absorbing UV line 
on the top. The horizontal bar indicates the observed band ratio (solid) 
and its 68\% uncertainties (dashed). \label{figA2}}
\end{figure}

\begin{figure}
\centering
\plotone{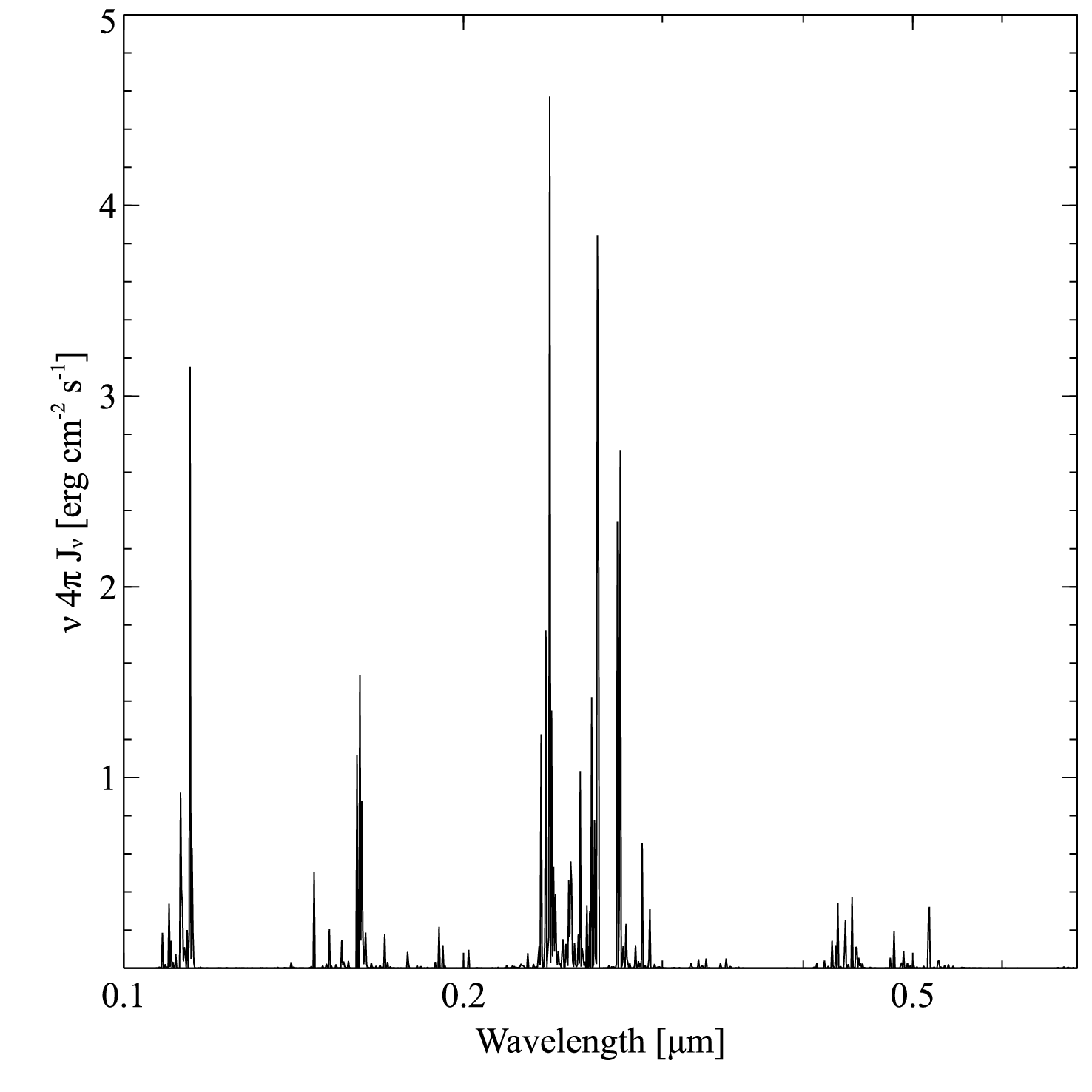}
\caption{The net Fe {\sc ii} emission from the cloud.  The 
horizontal axis gives the wavelength in microns and the vertical 
axis gives the emission produced by the cloud.  The incident 
continuum is not included since we assume that it is not directly 
visible to us. 
\label{figA4}}
\end{figure}

Results are shown in Figure \ref{figA2}. The lower axis gives the total 
hydrogen column density.  An iron abundances of log Fe/H = -4.55 was assumed.  
A good fit is obtained at $\log N(H) = 23$ cm$^{-2}$.  The upper axis shows 
the optical depth in one of the strong UV lines that is responsible for 
pumping the atom. The net \feii emission is shown in \ref{figA4}. \feii UV 
1, 3, 4, 63, 64 are the strongest lines, while those at wavelengths longer 
than 2800\AA are much weaker. The spectrum looks similar to the observed 
\feii. 

With these parameters, the thickness of the cloud is one parsec, and the 
distance to the continuum source is estimated about 1 kpc, following the 
definition of the ionization parameter, and the inferred intrinsic continuum 
luminosity of J1535+0903. The total volume of Fe {\sc ii} region is about 
$10^6$ parsec$^3$, and a total gas mass of about $10^9$ M$_\sun$, which is 
comparable to the reported $10^{8-9}$ M$_\sun$ molecular gas in outflows of 
infrared luminous quasars, such as Mrk 231 (Feruglio et al. 2010; Cicone et 
al. 2014). However, the gas density is poorly constrained so these numbers 
are only very rough estimate.   

\end{document}